# Charge Trapping Dynamics in PbS Colloidal Quantum Dot Photovoltaic Devices

Artem A. Bakulin[1*], Stefanie Neutzner[1], Huib J. Bakker[1], Laurent Ottaviani[2], Damien Barakel[2], Zhuoying Chen[3*]

[1] FOM institute AMOLF, Science Park 104, Amsterdam 1098 XG, The Netherlands

[2] Institut Matériaux Microélectronique Nanosciences de Provence, UMR CNRS 7334, Faculté des Sciences St Jérôme, Aix-Marseille Université, 13397 Marseille cedex 20, France

[3] Laboratoire de Physique et d'Etude des Matériaux, ESPCI/CNRS/UPMC UMR 8213, 10 rue Vauquelin, 75005 Paris, France

*Corresponding authors: a.bakulin@amolf.nl ; zhuoying.chen@espci.fr ;

**Abstract:**

The efficiency of solution-processed colloidal quantum dot (QD) based solar cells is limited by poor charge transport in the active layer of the device, which originates from multiple trapping sites provided by QD surface defects. We apply a recently developed ultrafast electro-optical technique, pump-push photocurrent spectroscopy, to elucidate the charge trapping dynamics in PbS colloidal-QD photovoltaic devices at working conditions. We show that IR photo-induced absorption of QD in the 0.2-0.5 eV region is partly associated with immobile charges, which can be optically de-trapped in our experiment. Using this absorption as a probe, we observe that the early trapping dynamics strongly depend on the nature of the ligands used for QD passivation while it depends only slightly on the nature of the electron-accepting layer. We find that weakly bound states, with a photon-activation energy of 0.2 eV, are populated instantaneously upon photoexcitation. This indicates that the photogenerated states show an intrinsically bound-state character, arguably similar to charge-transfer states formation in organic photovoltaic materials. Sequential population of deeper traps (activation energy 0.3-0.5 eV) is observed on the ~0.1-10 ns time scales, indicating that most of carrier trapping occurs only after substantial charge relaxation/transport. The reported study disentangles fundamentally different contributions to charge trapping dynamics in the nanocrystal-based optoelectronic devices and can serve as a useful tool for QD solar cell development.

Colloidal semiconducting quantum dots (QDs) have attracted extensive interest as active building blocks for low-cost solution-processed photovoltaics due to their size-tunable absorption from the visible to the near IR.[1] Advances in colloidal synthesis over the last two decades have enabled the production of high quality, II-VI and IV-VI (such as CdSe, CdS, PbSe and PbS) nanocrystal QDs of various sizes and morphologies.[2-5] Of these QD systems, the lead chalcogenide of PbSe and PbS, having a bulk bandgap of 0.28 eV and 0.41 eV respectively, are particularly attractive for photovoltaic applications due to their excellent photosensitivity in the near-IR. Over the last five years significant progress in the device performance of PbSe and PbS colloidal QD solar cells have been achieved by optimizing both device structures and nanocrystal surface ligands.[1,6-13] Recently a power conversion efficiency up to 8.5% has been demonstrated for a depleted $TiO_2$/PbS heterojunction structure that combines a donor-supply electrode strategy with the use of hybrid ligands for passivation.[14] Yet further improvements in device performance are still needed for this technology to become of commercial importance.[15] To achieve this goal a better understanding of material properties and charge trapping mechanisms is indispensable.

One of the factors limiting the efficiency of solution-processed colloidal QD solar cells is the inefficient charge extraction from the active layer of the device.[16] Recently, it was shown that the poor charge transport properties of QD films originates from the high concentration of defect states provided by the QD surface states.[12,17] Such defect states act as trapping centers for the photogenerated charges, decrease their charge mobility, enhance recombination, and thereby set a limit to the cell thickness and light-absorption efficiency.[8] For this reason, the control over the concentration and depth of surface traps has become a major tool for improving the photoconversion efficiency of QD devices. Such control is usually achieved by the management of ligands surrounding the QDs and atomic-level passivation of their surfaces.[6,8,12,18-20]

The dynamics of charge generation, trapping and recombination in QD thin films occurs on multiple timescales, being particularly rich in the first few ns after the excitation, as demonstrated by a large number of time-resolved spectroscopic studies that probed the carrier dynamics with near- and mid-IR light pulses. These studies make use of the fact that the electronic excitation of QDs leads to the appearance of new absorption bands in the near and mid-IR spectral regions.[8,17,21] The origin of this excitation-induced absorption is still debated, as it can be associated with mobile as well as trapped charge carriers or, most likely, represents a combination of responses from both sub-ensembles showing slightly different spectral characters and oscillator strengths. The generation of hot carriers occurs on a sub-ps time scale and is followed by ~ps intraband relaxation, ~100 ps Auger recombination and relatively slow >> ns geminate and bimolecular recombination of charges.[16,22-30] However, except for THz and microwave measurements that probe the local conductivity,[31] purely optical measurements are not capable of discriminating free from bound charges, nor do they provide a clear link between the spectroscopic observables and device performance. For these reasons the dynamics of the efficiency-limiting charge trapping process in QD devices are still not well understood.

Here we apply a combination of optical and electronic techniques to elucidate the charge dynamics in PbS colloidal QD photovoltaic devices at working conditions. Using pump-push photocurrent spectroscopy, that measures the photocurrent of the cell induced by mid-IR re-excitation, we selectively track in time the population of various trapped species and compare their dynamics to those of free charges. Our results show the presence of trapping sites of different origin and different binding energy that become populated at time scales ranging from 100 fs to 10 ns after the excitation.

**Results and Discussion**

QD photovoltaic devices were fabricated using a depleted heterojunction architecture (Figure 1a inset) with the junction formed between a nanocrystalline $TiO_2$ layer and a colloidal PbS QD film. As-synthesized QDs exhibited a diameter between 3 to 4 nm and were found to be well crystalline, as observed from annular dark-field scanning transmission electron microscopy (ADF-STEM) (Figure 1b). Thin films deposited from as-synthesized QDs exhibit their first excitonic transition at about 900 nm. The QD layer used in solar cells, about 150 nm in thickness, was deposited onto $TiO_2$ coated ITO substrates *via* a "layer-by-layer" spin-coating and ligand-exchange method.[32] Most of the results presented have involve exchanging the as-synthesized long-chain ligands (oleic acid) by 3-mercaptopropionic acid (MPA), a short-chain bifunctional ligand widely used in PbS QD solar cells.[8,33,34] FT-IR spectra on exchanged QD films reveal a significantly reduced intensity in the absorption peak around 2900 $cm^{-1}$ compared to non-exchanged films (Supporting information, Figure S1), which indicates the exchange of oleic acid ligands by MPA.[8] In order to perform comparison studies in pump-push photocurrent spectroscopy we also fabricated QD solar cells under identical conditions except using 1,2-Ethanedithiol (EDT) in ligand-exchange or ZnO as the acceptor layer. To improve hole extractions as well as electron-blocking[35] an interfacial layer of about 10-nm-thick $MoO_3$ was deposited on top of the QD layer, followed by a deposition of a 100-nm-thick gold contact. Subsequently, the device was encapsulated with resin and encapsulation glass in inert. The samples used in those experiments with a push energy <0.3eV were fabricated under identical conditions as described above except that a semi-transparent gold contact of about 8-nm-thick was used. As we could not apply the encapsulation for semi-transparent-electrode samples, the corresponding measurements were done under $N_2$ gas flow. The results for semi-transparent-electrode devices show good correspondence to the results obtained for thick-electrode encapsulated devices (figure S4, Supplementary information).

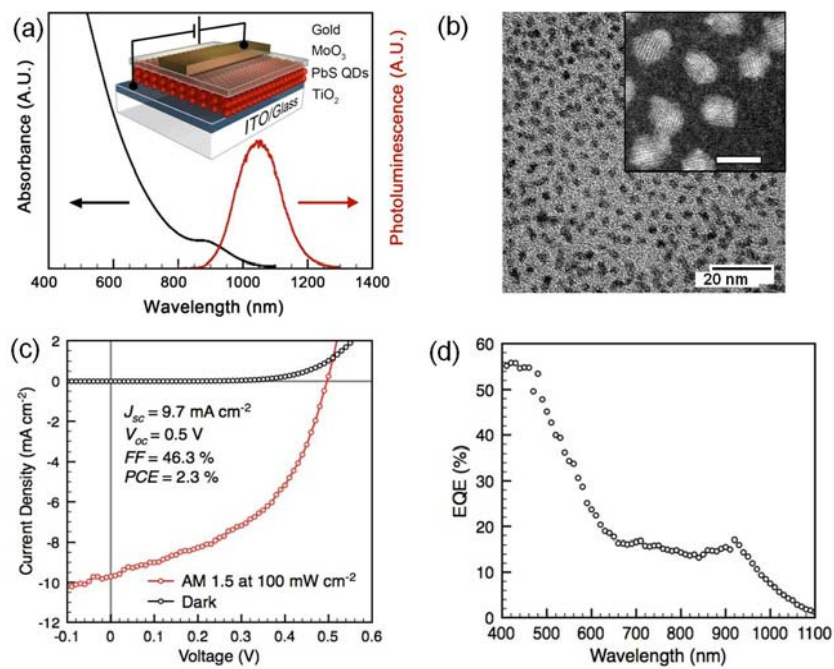

**Figure 1.** (a) UV-Vis-NIR absorbance and PL from PbS QD thin films (PL excitation at 515 nm). Inset: Schematic of the device architecture constituting the nanocystalline $TiO_2$ and PbS QD heterojunction; (b) Bright field TEM image of the PbS QDs. Inset: Annular dark-field scanning transmission electron microscopy (ADF-STEM) image revealing the crystallinity of each QD (inset scale bar represents 5 nm). (c) *J-V* characteristics of a representative $TiO_2$/PbS QD solar cell in the dark and under 100 mW cm$^{-2}$ AM 1.5 illumination. (d) The external quantum efficiency (EQE) spectra for a representative $TiO_2$/PbS QD solar cell solar cell.

Current-voltage (*J-V*) characteristics of the encapsulated solar cells were measured in the dark and under 100 mW·cm$^{-2}$ AM1.5 illumination. Representative *J-V* curves for $TiO_2$/QD-MPA depleted heterojunction solar cells are plotted in Figure 1c with the corresponding photovoltaic performance parameters. Averaged over 5 devices we obtained typically an open-circuit voltage ($V_{oc}$) of 0.46 ± 0.02 V, a short-circuit current ($J_{sc}$) of 10.4 ± 1 mA cm$^{-2}$, a fill-factor (*FF*) of 44 ± 4%, and an AM1.5 power conversion efficiency (PCE) of 2.1 ± 0.2%. Figure 1d shows the typical device external quantum efficiency (EQE), the ratio of extracted electrons to incident photons as a function of wavelength. The spectral response of such QD-based solar cells mainly rely on the photon absorption in the QD layer. Similar to the QD absorption profile, near 900 nm a distinctive excitonic feature is observed with an amplitude corresponding to about 17% in EQE.

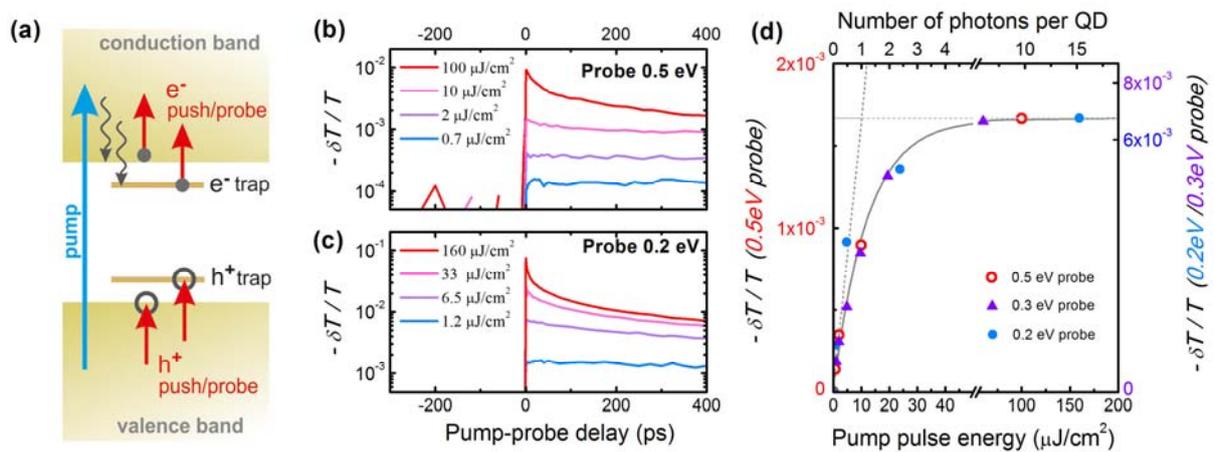

**Figure 2.** (a) The schematic diagram of the QD energy levels with optically-induced electron transitions indicated by arrows. The results of pump-probe experiments with different illumination fluxes performed on: (b) a QD photovoltaic device at short-circuit conditions with a probe photon energy of 0.5eV, (c) a QD film on a CaF$_2$ substrate with a probe photon energy of 0.2 eV. The pump photon energy was 1.8 eV. (d) The amplitude of the photoinduced absorption at 'long' 500ps delay as a function of pump intensity. The dashed curves show the limiting behavior at 'low' and 'high' photon fluxes. Their intersection provides the estimate for 'single exciton per QD' conditions.

Figure 2 presents the idea and the results of pump-probe experiments on a QD photovoltaic device with a probe photon energy of 0.5 eV and on a QD film with a probe photon energy of 0.2 eV (as an ITO electrode is not transparent in this probe region). The pump photon energy was 1.8 eV to avoid carrier multiplication effects.[36-39] The pump-probe transients demonstrate a 200-fs time resolution limited (see figure S6, Supplementary Information) build up of photoinduced absorption upon sample excitation. This indicates that our experiment probes intraband transitions which are well-known signatures of photogenerated charge carriers (both electron and holes are addressed).[22,23] However, from this result no conclusion about charge mobility can be drawn, as both trapped and free charges contribute to the signal in this probe energy region.[21] We observe a clear dependence of the signal kinetics on the illumination power. At low fluxes, when the probability of multiple excitation of the same QD is negligible, no decay is observed at the ps-ns time scale. As the flux increases, multiple decay components appear in the transients, which we assign to the relaxation of multi-exciton states to single-exciton states due to Auger recombination.[23,38] The dependence of the transients on the illumination power allows for an estimation of the average number of excitons per QD (figure 2d).[21] We note that a flux of 0.1 µJ/cm$^2$ (later used in the pump-push experiments) is well below the multi-exciton excitation regime (<0.05 excitons per QD). The lack of dynamics at positive pump-probe delays in the case of low illumination indicates that there is no sub-ns charge recombination occurring in the QD film. Interestingly, although the responses at 0.2 eV and 0.5 eV have different amplitudes (determined by the dipole strength), the signals show a very similar time-dependence, which suggests that similar sub-ensembles of electrons and holes are probed. However, this independence on the probe energy also indicates that the pump-probe signal is quite insensitive to charge transport and relaxation processes occurring in the QD sample.

We performed a detailed study of the character and dynamics of the trapped states in the QD layer using pump-push photocurrent spectroscopy (PPP), which is designed to probe the presence of bound charges in operational photovoltaic cells (figure 3a).[40-42] The operational device is exposed to a visible (~1.8 eV photon energy, 0.1 μJ/cm$^2$) pump pulse, which leads to the formation of trapped and free charge carriers. The photogenerated free carriers create a 'reference' photocurrent output $J$ of the cell. After a delay, the device is illuminated by an IR (0.2-0.5 eV, ~100 μJ/cm$^2$) push pulse which is absorbed by the generated electrons and holes,[8] thus providing them with extra energy. If the carriers are free, their dynamics are not influenced by the excess energy as they will quickly return by rapid thermalization to the state they were in before the excitation. In contrast, for charges that are confined in a low-energy trap state, the excess energy can lead to de-trapping, thereby providing the cell with an additional photocurrent $\delta J$. The normalized change in current $\delta J/J$ thus forms a measure of the relative amount of trapped states in the device. By measuring $\delta J/J$ as a function of the delay between the push and the pump pulses, we obtain information on the population and recombination kinetics of the trapped electrons and holes. We note that the push pulse by itself generates a noticeable current in the device. This current results from the direct excitation of sub-gap states of QD by the low-energy photons of the push pulse, and does not depend on the delay between the pump and push pulses (figure S5, Supplementary Information). This delay-independent background is subtracted in the data presented below.

Figure 3b shows the PPP kinetics of the QD cell measured for different push photon energies. The data were normalized to the reference current $J$~1nA and re-scaled, for the amplitudes to correspond to the same push pulse energy density of 100μJ/cm$^2$. The observed curves also contained a delay-independent background, probably related to the optical activation of trapped charges with >1ms lifetime and to direct excitation of QD sub-gap states; this background was subtracted to facilitate the comparison between different curves. The presented results clearly show that the cell photocurrent increases due to IR excitation. Therefore, the mid-IR absorption of the photo-induced charge carriers is, at least partly, associated with the excitation of trapped states. The IR excitation leads to de-trapping and thus increases the charge transport inside the device.

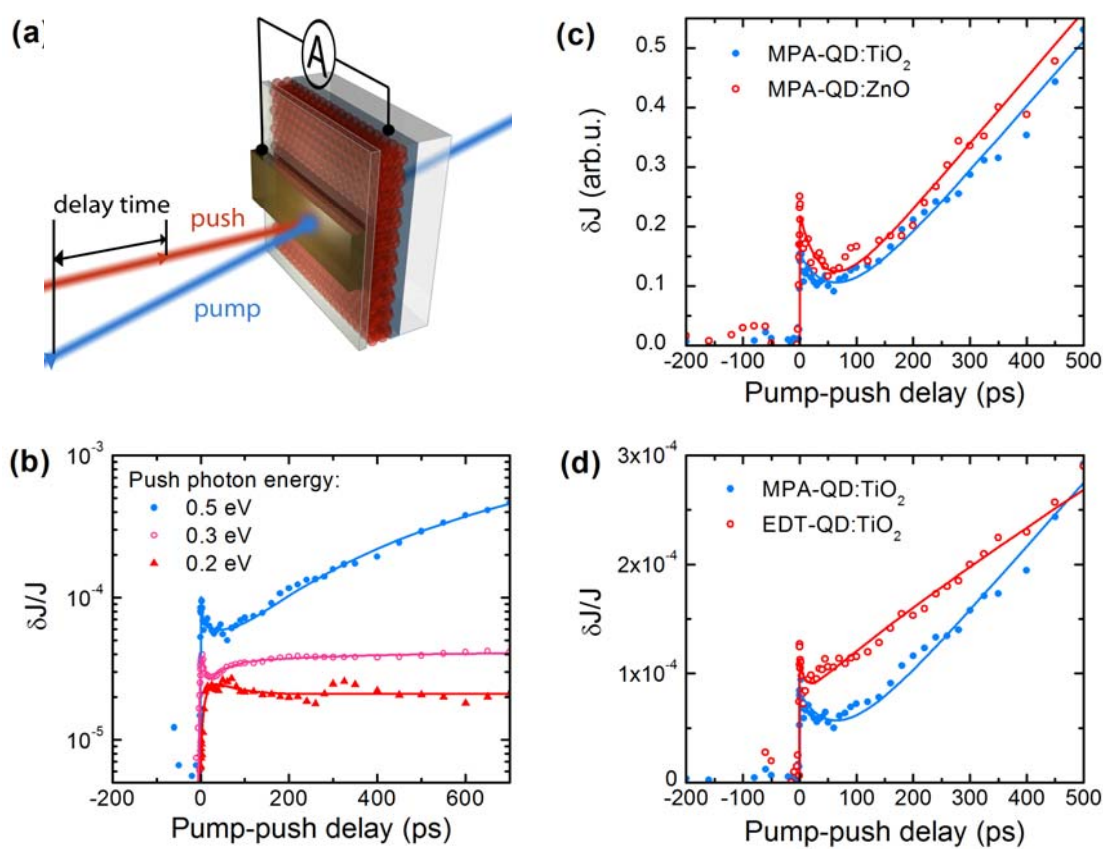

**Figure 3.** (a) The pump-push photocurrent experimental layout. (b) The results of PPP experiments on a QD device with different push photon energies (all signals correspond to a push beam energy density of 100 μJ/cm$^2$). (c) and (d) show PPP transients observed at a push photon energy of 0.5 eV for devices with different electron accepting layer and different QD ligands, respectively. All measurements were performed under short-circuit conditions with a pump photon energy of 1.8 eV. The data were corrected for the delay-independent background signal originating from the direct excitation of QD by push and long-lived trapped charges. The solid curves are multi-exponential fits to the data serving as guides to the eye.

All curves in figure 3b show a prompt initial growth followed by a minor decay, which reflects fast charge generation and, probably, early relaxation or recombination.[26] The immediate appearance of a PPP response indicates that the initial charge separation is not 100% efficient. The precise yield of trapped charges cannot be calculated as the efficiency of optical de-trapping is not known. However, assuming de-trapping with 0.5 eV photons to be 100% efficient, we can calculate the fraction of trapped states to be 0.2% after the excitation and 1% at longer times. The efficiency of optical activation is probably substantially lower and depends on the photon energy which means that there the fraction of trapped charges could be substantially higher than these estimated values.

The long time behavior of PPP response is dramatically different for different push photon energies, which indicates that the trap states that are activated by different photon energies have different population dynamics. Considering the fact that we did not observe substantial variations in the pump-probe transients at different mid-IR probe energies, the effect is unlikely to be associated with variations in the concentration of charge carriers or absorption cross-sections. A logical explanation for the difference in the PPP transients at different push-photon energies is that de-trapping happens from traps of different activation energy (depth). This idea is supported by previous studies[17] and also by the higher amplitude of the extra photocurrent induced by the higher energy push photons. At this point, it should be noted that the energy of the photon activating the trap may not reflect the charge binding potential directly, because the particular mechanism and the efficiency of the push-photon-induced de-trapping may depend strongly on the density and delocalization of states populated by the push pulse. We also note that at longer times the rise of the $\delta J/J$ signal will saturate and start to decay due to recombination of trapped charges, however, our observation time window is too short to observe this.

The increase of $\delta J/J$ for 0.3-0.5eV push photons occurs on relatively long time scales of ~0.1-10 ns after the initial charge generation. This time is much longer than the intraband relaxation of hot carriers and should be associated with charge transport in the device.[26] This indicates that the population of the deep traps is mediated by electron/hole diffusion process. The ns time scale is much shorter than the time needed for charges to arrive to the $TiO_2$:QD interface and to be extracted from QD layer.[43,44] This notion agrees with the fact that the PPP kinetics did not depend strongly on the type of electron-accepting layer used. This is illustrated by figure 3c which compares the PPP dynamics for devices with $TiO_2$ and ZnO acceptor layers. Therefore, push-induced activation of interfacial and oxide trapping states can be ruled out. The acceptor layer probably influences charge dynamics only 10-100 ns after the excitation, when diffusing electrons reach the QD:oxide interface.

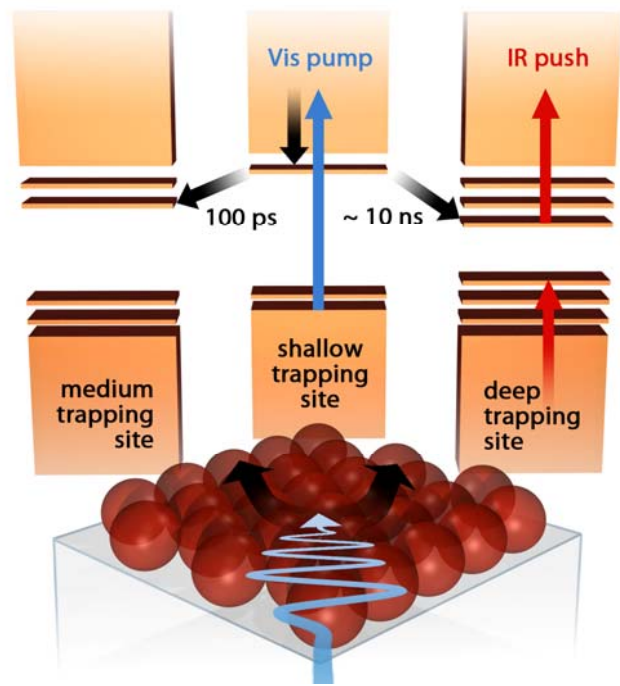

**Figure 4.** Schematic diagram illustrating charge trapping in QD solar cell. After photogeneration (transitions for electrons showed by arrows) charge carriers diffuse through the QD layer they are sampling trapping states of different energy. The deepest traps with an energy of ~0.5 eV are populated a few ns after the initial charge photogeneration.

Figure 4 presents a model for carrier trapping that emerges from our observations. QDs in the active layer have a wide distribution of trapping sites. Some of these sites are populated almost immediately after the photo-excitation. During the transport to the electrodes, the mobile carriers diffuse through the QD layer, sampling on the way trapping states of different energy. The traps which can be activated with ~0.3 eV photons are mostly populated within a few 100 ps. The deepest traps (~0.5eV activation) are populated ~10 ns after the photo-excitation of charges. These deeper traps are most likely responsible for the low device performance. The sensitivity of the observed dynamics to the active layer properties is illustrated in figure 3d which compares the PPP dynamics for cells fabricated with QDs exchanged with different ligands. The charge trapping dynamics in devices using MPA and EDT ligands is substantially different, which can be used as contrast parameter for cell optimization.[8]

Interestingly, at each photon energy there is a sub-ensemble of trapped carriers that respond to the push pulse instantaneously (figure 3b), indicating that part of the carriers is immediately photo-generated in bound states that do not contribute to the photocurrent. The electrons and holes in QDs are linked together by the excitonic binding energy. Moreover, the binding may be enhanced due to the effect of surface defects on the thermalisation dynamics and the localization of carriers.[26] At 0.2 eV this is the only sub-ensemble and although no recombination occurs in the material, as seen in pump-probe data, the amount of trapped charges at this energy does not grow with time. Such behavior is very similar to organic[45-50] and hybrid[51-55] photovoltaic devices, where the majority of immobile charges are formed almost immediately after excitation, in a so-called bound charge-transfer state. The effects of carrier-carrier interactions on their mobility have been observed before in inorganic semiconductors,[56] however, except for interface-related injection effects,[44,57-59] such phenomena have rarely been discussed as a loss mechanism in nanocrystal-based devices. On the opposite, the charge generation in strongly coupled QD films approaches unity.[31] Our experiments now indicate that the initial long-range separation of photoinduced charge carriers in working devices may not be 100% efficient. The seeming controversy with the previous works on PbSe QDs [31] [36] can be explained by the higher dielectric constant of PbSe,[60] the effect of carrier multiplication in PbSe QDs, and/or the effect of incomplete surface passivation which is more difficult to control inside the device active layer. In addition, during the transport to electrodes the charge carriers that are initially mobile become to a large extent trapped in deep traps, thereby further limiting the device performance.

## Conclusions

We applied a combination of novel ultrafast electrical-optical techniques to elucidate the charge trapping dynamics in different colloidal PbS QD-based photovoltaic devices at working conditions. Our results demonstrate the presence of different mechanisms of charge immobilization in QDs, coming into play on diverse timescales after the photo-induced carrier generation. We observe that the initial charge separation is not 100% efficient, and that the overall device performance is further limited by carriers falling in deep (up to 0.5eV) traps during charge transport. These observations provide new insights into the fundamental physical mechanisms that limit the performance of QD-based photovoltaic devices. The newly developed electrical-optical technique also shows sensitivity to the trapping dynamics of different material systems and thus forms a useful tool for photovoltaic cell optimization.

## Methods

**Synthesis of PbS QDs and characterizations:** PbS QDs were prepared by a procedure modified from the method developed by Hines *et al.*[2] For the formation of lead oleate, in a 50 mL three-neck flask, 18 mL octadecene, 0.45 g lead (II) oxide, and 1.5 mL oleic acid were stirred and degassed under vacuum at 100°C until lead oxide was completely dissolved. This mixture was then heated to 125°C under argon flow. A sulfur precursor was prepared separately inside a glovebox by mixing 10 mL of octadecene (degassed previously) and 0.18 mL of hexamethyldisilathiane. This sulfur precursor was then injected into the lead oleate solution at 125°C. After injection the heating was switched off and the reaction mixture was allowed to cool down to 40°C in about 40 minutes. Acetone was then added to the reaction mixture to precipitate the QDs by centrifugation. The precipitate was then dispersed in toluene, and re-precipitated one more time by acetone and centrifugation. The precipitate was then re-dispersed in toluene and processed further by at least two additional precipitations by adding a mixture of methanol and butanol. The final precipitate was transferred inside an argon-filled glovebox and re-dispersed in anhydrous octane. UV-Visible absorption spectra on thin film samples were measured in air using a Lambda 25 UV/Vis spectrometer. Thin film photoluminescence measurements were carried out at room temperature with a cryostat (vacuum <$10^{-5}$ mbar) by a Horiba Jobin Yvon PL system with a liguid nitrogen cooled InGaAs detector and an argon-ion laser (excitation wavelength: 515 nm). All film thicknesses were measured using a profilometer (Veeco Dektak).

**Device fabrication and characterization:** All solvents were used as purchased without further purification. Indium-tin oxide (ITO) substrates were first cleaned in an ultrasonic bath of 10% NaOH solution followed by a rinsing step with deionized water. The substrates were then further sonicated in a bath of deionized water, acetone, and isopropanol. A $TiO_2$ nanoparticle suspension (Solaronix HT-L/SC) was spin-coated on the substrates followed by annealing in air at 210 °C for 60 minutes to form a nanocrystalline $TiO_2$ layer of about 100 nm in thickness. For solar cells that used ZnO as the acceptor, the ZnO layer was prepared by a method reported previously.[61,62] Typically 1 g of zinc acetate dihydrate and 0.28 g of ethanolamine were dissolved in 10 mL of 2-methoxyethanol under vigorous stirring for 12 hours in air. This ZnO-precursor solution was then spun onto cleaned ITO substrates, followed by thermal annealing in air at 200 °C for 1 hour. The TiO$_2$ or ZnO coated ITO substrates were then transferred to an argon-filled glovebox for a layer-by-layer QD spin-coating process.[32] After spin-coating of each sub-layer, we dipped the sample into a solution of acetonitrile containing 10% 3-mercaptopropionic acid (MPA) or 1,2-ethanedithiol (EDT) for 30s followed by a rinse with clean acetonitrile. The effect of ligand exchange was observed with FTIR spectroscopy (figure S1, Supplementary information). We repeated the layer-by-layer spin-coating 12 times until the film thickness reaches about 140 nm. A 10-nm-thick MoO$_3$ interfacial layer was then thermally evaporated in vacuum ($< 4\times10^{-6}$ mbar) through a shadow mask onto the QD layer, followed by evaporation of a 100-nm-thick gold contact. The devices were then encapsulated under inert conditions before testing them in air. The devices used in the pump-probe and the pump-push experiments at 0.5 eV probe/push photon energy were the same as described above. For the experiments in which push-photon energies of 0.2-0.3 eV were used, another batch of devices was fabricated under identical conditions as described above except that a semitransparent gold contact of about 8-nm-thick was used and no encapsulation was applied.

The current-voltage characteristics of the solar cells were measured using a Keithley 4200-SCS Semiconductor Characterization System. The devices were illuminated through the glass substrate using an Oriel 91160-1000 full spectrum solar simulator with AM1.5 G filters. The mismatch between the simulator and solar spectra was not corrected for during these measurements. For the EQE measurements, we used a monochromatic light beam obtained from a white light source and a monochromator (and appropriate filters to avoid high-order harmonics). The light beam was chopped at 45 Hz. With an optical fiber, a diaphragm, an inverted optical microscope (NachetMS98) and associated objective lens we obtained a monochromatic light spot on our devices (spot diameter about 1.5 mm and light intensity < 1 mW cm$^{-2}$). The monochromatic light intensity was calibrated by a NIST-calibrated Si diode. The short-circuit current was amplified by a low-noise current preamplifier (SR570) and then measured by a lock-in amplifier (SR810).

**Pump-Push Photocurrent Spectroscopy:** The output of a regenerative 1 kHz Ti:Sapphire amplifier system (Coherent, Legend Elite Duo, 800 nm, 40 fs pulse duration, 7 mJ per pulse) was split into two parts. One part was used to pump a broadband non-collinear optical amplifier (Clark) to generate visible pump pulses (100 fs pulse duration, 1.8 eV photon energy). Another part was used to generate mid-IR probe/push pulses by pumping either a 3-stage home-built optical parametric amplifier (100 fs pulse duration, 0.5 eV photon energy), or commercial parametric amplifier with a difference frequency generation stage (HE TOPAS, 150 fs, 0.2 eV and 0.3 eV photons).

In the pump-push photocurrent experiments all devices were measured at short-circuit conditions. ~1 nJ pump and ~1 μJ push pulses were focused onto a ~1 mm$^2$ spot on the device through a semitransparent gold top electrode. The reference photocurrent induced in the studied device by the pump was detected at the laser repetition frequency of 1 kHz by a lock-in amplifier. The push beam was mechanically modulated at ~380 Hz, and its effect on the photocurrent was detected by a lock-in amplifier locked to the chopper frequency. The polarization of pump beam was set by an achromatic half-wave plate and thin-film polarizer (1:200 extinction) 54 degrees (magic angle) with respect to the polarization of the push beam. To avoid experimental artifacts like multi-photon contributions, we measured the intensity dependence of the signal. For 0.5 eV push photons, pump and push irradiated the active layer through the ITO electrode which is (semi)transparent in this spectral region. We also repeated the experiment on the cell with thick gold electrode instead of semitransparent one (figure s4, Supplementary Information). The experiments with 0.2 eV and 0.3 eV push photons were performed only with semitransparent gold electrode devices as ITO absorbs in this region. To avoid sample degradation the measurements were performed under nitrogen gas flow.

**Pump-probe spectroscopy:** Pump-probe transients for a 0.5 eV photon probe were measured on the device with thick gold back electrodes. After passing the active layer, IR probe beam was reflected by the gold electrode back and detected by a commercial PbS photodiode. The pump-probe transients with the 0.2 eV probe were measured on a QD film (identical to active layer of the device) deposited on top of $CaF_2$ substrate. The probe and reference IR beams passed through the film and were detected by a nitrogen-cooled MCT detector array. The measurements were performed under $N_2$ flow to avoid water vapor absorption of IR light and sample degradation.

**Acknowledgements**

A.A.B thanks Erik Garnett, Akshay Rao and Brian Walker for stimulating discussions. Z.C. is grateful to Martiane Cabié, Thomas Neisius, Guillaume Radtke, and Xiangzhen Xu for their assistance in TEM and STEM characterizations. A.A.B. acknowledges a VENI grant from the Netherlands Organization for Scientific Research (NWO). Z.C. acknowledges support from the ANR-2011-JS09-004-01-PvCoNano project and the EU Marie Curie Career Integration Grants (303824).

**Supporting Information**

FTIR spectra of QD films before and after ligand exchange, IV curve of the $TiO_2$:QD-EDT device, IV curve of the ZnO:QD-MPA device, comparison of the PPP transients measured for devices with a semi-transparent and with a thick back electrode, PPP results without background subtraction, short-time pump-probe dynamics, and IR-induced current as a function of light intensity plot. This material is available free of charge *via* the Internet at http://pubs.acs.org.

**TOC graphics:**

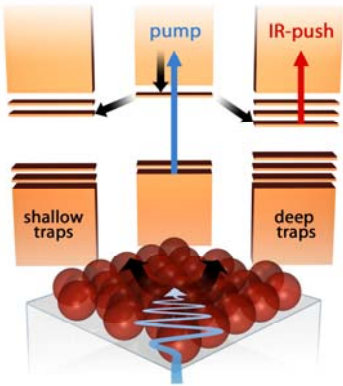 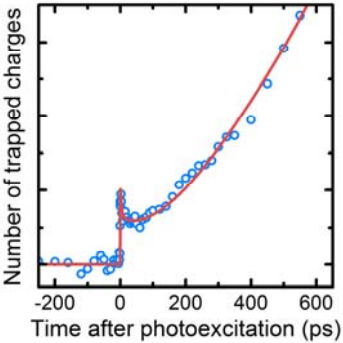